\documentclass[9pt]{article}
\usepackage{graphicx}
\usepackage{epsfig}
\usepackage{subfigure}  
\usepackage{float}
\usepackage{braket}
\usepackage{pslatex,latexsym,times}
\usepackage{bm, float, lipsum}
\usepackage{amsmath}
\usepackage{amssymb}
\usepackage{amsfonts}
\usepackage{mathrsfs}
\usepackage{color}
\usepackage{rotating}
\usepackage[normalem]{ulem}
\usepackage[labelfont=bf]{caption}

\begin{document}
\title{Optimal Localization of Diffusion Sources in Complex Networks}

\author
{Zhao-Long Hu,${}^{1}$  Xiao Han,${}^{1}$ Ying-Cheng Lai,${}^{2,4}$  Wen-Xu Wang,${}^{1,2,3}$\\
\\
\normalsize{${}^{1}$School of Systems Science, Beijing Normal University, Beijing, 100875, P. R. China}\\
\normalsize{${}^{2}$School of Electrical, Computer and Energy Engineering, Arizona State University,}\\
\normalsize{Tempe, Arizona 85287, USA}\\
\normalsize{${}^{3}$Business School, University of Shanghai for Science and Technology, Shanghai 200093, China}\\
\normalsize{${}^{4}$Department of Physics, Arizona State University, Tempe, Arizona 85287, USA}\\
\\
\normalsize{E-mail:  wenxuwang@bnu.edu.cn}
}

\date{}
\maketitle
\begin{abstract}
Locating sources of diffusion and spreading from minimum data is a significant problem in network science with great applied values to the society. However, a general theoretical framework dealing with optimal source localization is lacking. Combining the controllability theory for complex networks and compressive sensing, we develop a framework with high efficiency and robustness for optimal source localization in arbitrary weighted networks with arbitrary distribution of sources. We offer a minimum output analysis to quantify the source locatability through a minimal number of messenger nodes that produce sufficient measurement for fully locating the sources. When the minimum messenger nodes are discerned, the problem of optimal source localization becomes one of sparse signal reconstruction, which can be solved using compressive sensing. Application of our framework to model and empirical networks demonstrates that sources in homogeneous and denser networks are more readily to be located. A surprising finding is that, for a connected undirected network with random link weights and weak noise, a single messenger node is sufficient for locating any number of sources. The framework deepens our understanding of the network source localization problem and offers efficient tools with broad applications.
\end{abstract}

\section{Introduction}
Dynamical processes taking place in complex networks are ubiquitous in natural and in technological systems~\cite{vespignani}, examples of which include disease
or epidemic spreading in the human society~\cite{neumann2009emergence,chin2013despite}, virus invasion in computer and mobile phone
networks~\cite{MayInternet,wang2009understanding}, behavior
propagation in online social networks~\cite{onlinespread}
and air or water pollution diffusion~\cite{pope2002lung,shao2006city}. Once an epidemic or environmental pollution emerges,
it is often of great interest to be able to identify its source within the network
accurately and quickly so that proper control strategies can be devised
to contain or even to eliminate the spreading process. In general,
various types of spreading dynamics can be regarded as diffusion
processes in complex networks, and it is of fundamental interest to be
able to locate the {\em sources of diffusion}. A straightforward,
brute-force search for the sources requires accessibility of global
information about the dynamical states of the network. However, for large
networks, a practical challenge is that our ability to obtain and process
global information can often be quite limited, making brute-force search
impractical with undesired or even disastrous consequences. For example,
the standard breadth-first search algorithm for finding the shortest paths,
when being implemented in online social networks, can induce information
explosion even for a small number of searching steps~\cite{broder2000graph}.
Recently, in order to locate the source of the outbreak of Ebola virus in
Africa, five medical practitioners lost their lives~\cite{gire2014genomic}.
All these call for the development of efficient methodologies to locate
diffusion sources based only on limited, practically available information
without the need of acquiring global information about the dynamical states
of the entire network.

There were pioneering efforts in addressing the source localization
problem in complex networks, such as those based on the maximum likelihood
estimation~\cite{source-pinto}, belief
propagation~\cite{altarelli2014bayesian}, the phenomena of hidden
geometry of contagion~\cite{brockmann2013hidden}, and inverse
spreading~\cite{zhu2013information,shen2015locating}. In addition, some
approaches have been developed for identifying super spreaders that
promote spreading processes stemming from
sources~\cite{Kitsak2010Identification,Pei2014Searching,Morone2015Influence}.
In spite of these efforts, achieving accurate source localization
from a small number of measurements remains challenging.
Prior to our work, a systematic framework dealing with the localization
of diffusion sources for arbitrary network structures and interaction
strength was missing.

In this paper, we develop a theoretical framework to address the
problem of network source localization in a detailed and
comprehensive way. The main focus is on the fundamental issue of
{\em locatability}, i.e., given a complex network and limited (sparse)
observation, are diffusion sources locatable? A practical and extremely
challenging issue is, given a network, can a minimum set of nodes be
identified which produce sufficient observation so that
sources at arbitrary locations in the network can actually be located?
To address these issues in a systematic manner, we use a two-step
solution strategy. First, we develop a minimum output analysis to identify
the minimum number of messenger/sensor nodes, denoted as $N_\text{m}$,
to fully locate any number of sources in an efficient way. The ratio of
$N_\text{m}$ to the network size $N$, $n_\text{m}\equiv N_\text{m}/N$,
thus characterizes the source locatability of the network in the sense
that networks requiring smaller values of $n_\text{m}$ are deemed to have
a stronger locatability of sources. Our success in offering the minimum
output analysis stems from taking advantage of the dual relation between the
recently developed controllability theory~\cite{yuan2013exact} and the
canonical observability theory~\cite{kalman1959general}. Secondly, given
$N_\text{m}$ messenger nodes, we formulate the source localization problem
as a sparse signal reconstruction problem, which can be solved by using
compressive sensing (CS)~\cite{candes2006robust,donoho2006compressed}, a
convex optimization paradigm. The basic properties of CS allow us to
accurately locate sources from a small amount of measurement
from the messenger nodes, much less than that required in the conventional
observability theory. We use our framework to examine a variety of model
and real-world networks, and offer analytical prediction of $n_\text{m}$
and demonstrate good agreement with numerical calculations. We find that
the connection density and degree distribution play a significant role in
source locatability, and sources in a homogeneous and denser network are
more readily to be located, which differs from existing algorithms for
source localization in the literature~\cite{source-pinto,zhu2013information,
shen2015locating}. A striking and counterintuitive finding is that,
for an undirected network with one connected component and random link
weights, a single messenger node is sufficient to locate any number
of sources in the presence of weak noise.

Theoretically, the combination of the minimum output analysis (derived from
the controllability and observability theories for complex networks) and
the CS-based localization method constitutes a general framework for
locating diffusion sources in complex networks. It represents a
powerful paradigm to exactly quantify the source locatability of a
network and to actually locate the sources efficiently and accurately.
Because of the CS-based methodology, our framework is robust against
noise~\cite{cande2008introduction,wang2011predicting},
paving way to practical implementation in noise environment.

\section{Results}
\subsection{A general framework to locate sources with minimum number of
messenger nodes}
\begin{figure}[!htb]
\begin{center}
\epsfig{figure=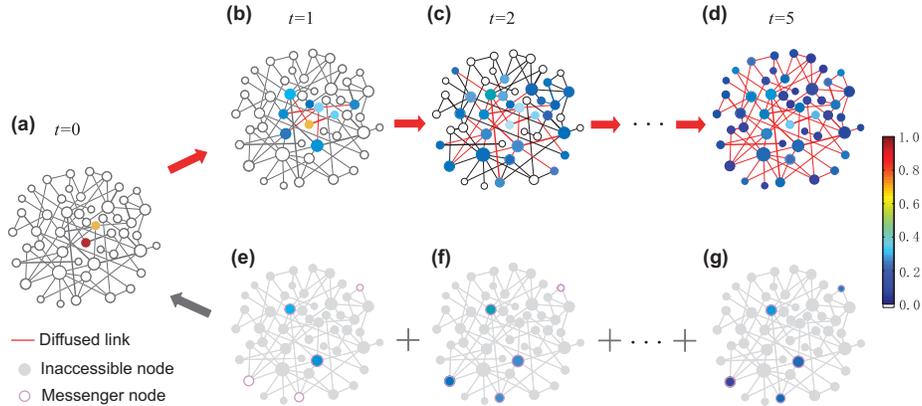,width=\linewidth}
\caption{Illustration of source localization problem.
(a) A random network with two sources at the initial time $t=0$. (b-d) The diffusion process at $t=1$ (b), $t=2$ (c) and $t=5$ (d), respectively. The color bar represents the state of node $x_i(t)$ and those links, along which diffusion occurred, are marked with red. (a) to (d) describe a diffusion (spreading) process from two sources to the whole network according to Eq.~(\ref{eq:model}). (e-g) Five messenger nodes whose states at three time constant can be measured and collected. The messenger nodes are specified by the output matrix $C$ and the states of messenger nodes and inaccessible nodes constitute ${\bf y}(t)$. The time of (e), (f) and (g) corresponds to (b), (c) and (d), respectively. However, in the real situation, the time as well as the initial time is unknown. The only available information for locating sources is the states of a set of messenger nodes at some time and the network structure. (e), (f) and (g) to (a) describe the source localization problem to be solved. Moreover, we aim to identify a minimum set of messenger nodes to locate arbitrary number of sources at any location by virtue of our minimum output analysis and optimization based on compressive sensing.
}
\label{Fig.0}
\end{center}
\end{figure}
We consider a class of diffusive processes on
networks, described by
\begin{equation} \label{eq:model}
x_{i}(t+1)=x_{i}(t)+\beta \sum_{j=1}^{N}\left[w_{ij}x_{j}(t)-w_{ji}x_{i}(t)\right].
\end{equation}
This equation
constitutes a good approximation for different types of linear diffusion processes and the linearization of some nonlinear diffusion processes~\cite{gomez2013diffusion}. For example, epidemics can be treated as linear dynamics in the early stages if the network connectivity is high.
Variable $x_{i}(t)$ that denotes the state of node $i$ at time $t$ captures the fraction of infected individuals, the concentration of
water or air pollutant and etc. at place $i$.
$\beta$ is the diffusion coefficient, $w_{ij}$ ($w_{ji}$) is the weight
of the directed link from node $j$ to node $i$ ($i$ to $j$), ($w_{ij}=w_{ji}$
for undirected networks), and $N$ is the number of nodes in the network (size). It is noteworthy that the value of the diffusion parameter $\beta$ should be constrained to ensure the physical meaning of $x_i(t)$, i.e., $x_i(t)$ is confined in the range $[0, 1]$ at any time $t$ for any node. We can prove that the confine of $x_i(t)$ leads to
$\beta\in(0,\min_{i=1,2,\cdots,N}\frac{1}{\sum_{j=1,j\neq i}^{N} w_{ji}}]$ (see Supplemental Material S1 for the proof).
Equation~(\ref{eq:model}) is discrete in time, facilitating greatly
computation and analysis. When observations are made from a subset of
nodes - the messenger nodes, system~(\ref{eq:model}) incorporating
outputs from these nodes can be written concisely as
\begin{equation} \label{eq:model_messenger}
\begin{cases}
{\bf x}(t+1)=(I+\beta L){\bf x}(t), \\
{\bf y}(t)=C{\bf x}(t),
\end{cases}
\end{equation}
where ${\bf x}(t)\in\mathbb{R}^{N}$ is the state vector of the entire
network at time $t$, $I\in\mathbb{R}^{N\times N}$ is the identity matrix,
$L=(W-D)$ is a Laplacian matrix, $W\in\mathbb{R}^{N\times N}$ is the weighted
adjacency matrix of elements $w_{ij}$, $D\in\mathbb{R}^{N\times N}$ is a diagonal
matrix of elements $d_{i}$ denoting the total out-weight
$\sum_{j\in\Gamma_i}w_{ji}$ of node $i$, where $\Gamma_i$ is the neighboring
set of $i$. The vector ${\bf y}(t)\in\mathbb{R}^{q}$ is the output at
time $t$ and $C\in\mathbb{R}^{q\times N}$ is the {\em output matrix}.
Messenger nodes are specified through matrix $C$ and ${\bf y}(t)$
records the states of these nodes. The source localization problem is
illustrated in Fig.~\ref{Fig.0}, which is a kind of inverse problem for diffusion and
spreading dynamics on complex networks.

The basic difference between source nodes and other nodes in the network
is that initially ($t = t_0$), the states of the former are nonzero
while those of the latter are zero. To achieve accurate localization of
an arbitrary number of sources at arbitrary locations, it is only necessary
to recover the initial states of all nodes from the measurements of the
messenger nodes at a later time ($t > t_0$). A solution to this problem
can be obtained using the observability condition in canonical control
theory. To be specific, we consider instants of time:
$t_0,~t_1,~\cdots ,~t$, and perform a simple
iterative process that yields the relation between ${\bf x}(t)$ and
${\bf x}(t_0)$: ${\bf x}(t)=\big[I+\beta L\big]^{t-t_{0}}{\bf x}(t_{0})$.
Consequently, the output, which depends on ${\bf x}(t_{0})$, can be
expressed as ${\bf y}(t)=C(I+\beta L)^{t-t_{0}}{\bf x}(t_{0})$.
The key to accurate localization of sources lies in the existence of a
unique solution of the equation, given the output vector
${\bf y}(t)$ from the set of messenger nodes as specified by $C$.
Intuitively, to obtain a unique solution, no fewer than $N$ snapshots
of measurement are needed. Without loss of generality, we assume that
uninterrupted time series from $t_0$ to $t_0+N-1$ are available. We obtain
\begin{eqnarray} \label{eq:main}
{\bf Y} = O \cdot {\bf x}(t_{0}),
\end{eqnarray}
where ${\bf Y} \in \mathbb{R}^{qN}$, the initial state vector is
${\bf x}(t_0) \in \mathbb{R}^{N}$, $q$ is the number of messenger nodes,
and the matrix $O \in \mathbb{R}^{qN\times N}$ is nothing but the
observability matrix in the canonical control theory (see Methods 5.1
for details of Eq.~(\ref{eq:main})). The observability full rank
condition~\cite{kalman1963mathematical} stipulates that, if and only if
${\rm rank}(O)=N$, there exists a unique solution of Eq.~(\ref{eq:main})
and the state vector ${\bf x}(t_{0})$ at initial time $t_0$ is observable.
Insofar as the given output matrix $C$ satisfies the observability rank
condition, the initial states of the nodes can be fully reconstructed
from the states of the messenger nodes, and all sources can then be
located. A challenge is that, in a realistic situation, the initial time
$t_0$ is often unknown, rendering the immediate application of the canonical
observability condition invalid. However, a unique and desired feature of
our framework is that both ${\bf x}(t_{0})$ and $t_0$ can be inferred based
on CS (see subsection {\it source localization} and Methods 5.2).
Thus, it is possible to develop a theoretical framework on the basis of the
observability condition (see Supplemental Material S2 for continuous-time processes).


\subsection{Minimum number of messengers for source localization}
Beyond the canonical observability theory, here our goal is to identify a
minimum set of messenger nodes to satisfy the
full rank condition for observability. However, the brute-force method
of enumerating all possible choices of the messenger nodes is
computationally prohibitive~\cite{liu2013observability}, as the total
number of possible configurations is $2^N$. Our solution is to use the
recently developed, exact controllability framework~\cite{yuan2013exact}
based on the standard Popov-Belevitch-Hautus (PBH) test
theory~\cite{hautus1969controllability}
and to exploit the dual relationship between controllability and
observability~\cite{kalman1959general}, which results in a practical
framework to find the required $N_\text{m}$ messenger nodes.
In particular, for an arbitrary network, according to the PBH test and the
exact controllability framework, $N_{\rm m}$ is determined by the maximum
geometric multiplicity of the eigenvalues $\lambda_i$ of the matrix
$I+\beta L$. After some matrix calculation, we obtain that (see Supplemental Material S3)
\begin{equation} \label{eq:MMT}
N_\text{m} = \max_{i}\{N-{\rm rank}[\lambda_{i}^{L}I-L]\},
\end{equation}
where $\lambda_i^{L}$ is the eigenvalue of matrix $L$ and
$\mu(\lambda_i^L) \equiv N-{\rm rank}[\lambda_{i}^{L}I-L]$ is the
geometric multiplicity of $\lambda_i^{L}$.
It is worth noting that the formula of $N_\text{m}$ does not contain
the diffusion parameter $\beta$, indicating that choices of $\beta$
do not affect the locatability measure $n_\text{m}$.
Equation~(\ref{eq:MMT}) as a result of the standard PBH test is a general
minimum output analysis for arbitrary networks.


For an undirected network, $L$ is symmetric and the geometric multiplicity
is nothing but the eigenvalue degeneracy. In addition, the eigenvalue
degeneracy of $L$ is equal to that of $I+\beta L$ (see Supplemental Material S3). Thus,
$N_{\rm m}$ is determined by the maximum eigenvalue degeneracy of $L$ as
\begin{equation} \label{eq:MMT2}
N^\text{undirect}_\text{m} = \max_{i}\{\delta(\lambda_i^L)\},
\end{equation}
where $\delta(\lambda_i^L)$
is the degeneracy of $\lambda_i^L$ (the number of appearances of
$\lambda_i^L$ in the eigenvalue spectrum). Equation~(\ref{eq:MMT2})
based on the PBH test is our minimum output analysis for arbitrary
undirected networks.

Equations~(\ref{eq:MMT}) and (\ref{eq:MMT2}) are the exact theory (ET)
for minimum output $N_\text{m}$ without any approximations, but the
associated computational cost resulting from calculating the eigenvalues
and identifying maximum value through a large number of comparisons in
Eqs.~(\ref{eq:MMT}) and (\ref{eq:MMT2}) is generally high.
Taking advantage of the ubiquitous sparsity of real
networks~\cite{strogatz2001exploring}, we can obtain an alternative
method to estimate $N_\text{m}$ with much higher efficiency. In particular,
for sparse networks, we have (see Supplemental Material S4):
\begin{equation} \label{eq:FA}
n^\text{sparse}_{\rm m} \approx 1-\frac{{\rm rank}(aI-L)}{N},
\end{equation}
where for undirected networks, $a$ is either zero or the diagonal element
with the maximum multiplicity (number of appearances in the diagonal) of
matrix $L$.
The matrix rank as well as eigenvalues in formula~(\ref{eq:FA}) can be computed
using fast algorithms from computational linear algebra, such as
SVD with the computation complexity $O(N^3)$~\cite{golub2012matrix} or LU decomposition with
the computation complexity $O(N^{2.376})$~\cite{cormen2001introduction}.
In general, Eq.~(\ref{eq:FA}) allows us to compute $n_\text{m}$ efficiently,
thereby the term {\em fast estimation} (FE) method.





\subsection{Analytical results for model networks}
\begin{figure}[!htb]
\begin{center}
\epsfig{figure=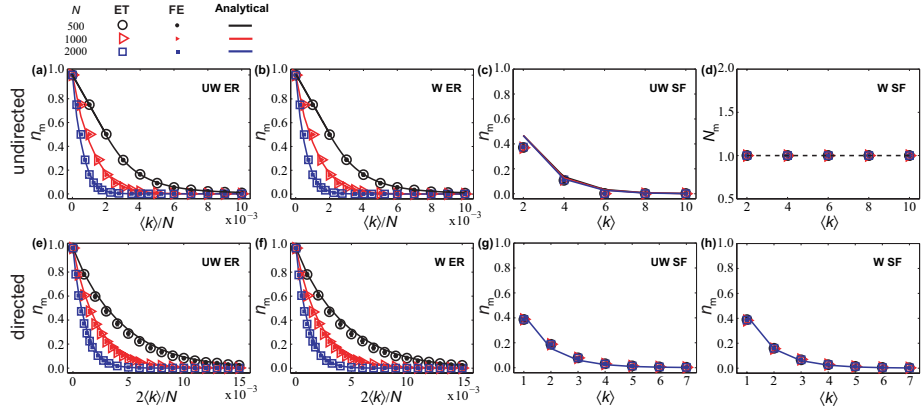,width=\linewidth}
\caption{Locatability measure $n_{\rm m}$ for ER and SF networks.
(a-b) For undirected networks, source locatability measure $n_{\rm m}$ as
a function of the connecting probability $\langle k\rangle /N$ for (a) unweighted ER
networks and (b) weighted ER networks. (c-d) $n_{\rm m}$ as a function of the average degree $\langle k\rangle$ for (c) unweighted SF networks, and
$N_{\rm m}$ as a function of the average degree $\langle k\rangle$ for (d) weighted SF networks.
For undirected networks, the values of $n_\text{m}$ are obtained from
the exact theory [ET; Eq.~(\ref{eq:MMT2})], fast estimation [FE;
(\ref{eq:FA})], and analytical prediction
(Analytical), for different network sizes. The analytical prediction
for ER networks is based on Eq.~(\ref{eq:UER}). For SF networks
in (c), the prediction is from the cavity method. (e-h) For directed networks, source locatability measure $n_{\rm m}$ as a function of the connecting probability
$2\langle k\rangle/N$ for (e) unweighted and (f) weighted ER networks, and as a
function of $\langle k\rangle$ for (g) unweighted and (h) weighted SF networks.
For directed networks, the ET results come from Eq.~(\ref{eq:MMT}), while
the FE results for ER and SF networks are from Eq.~(\ref{eq:FA}). The analytical predictions for
ER and SF networks are from Eq.~(\ref{eq:DER}) and
Eq.~(\ref{eq:DSF}), respectively. For weighted networks, link weights
are randomly selected from a uniform distribution in the range $(0,~2)$, which leads to that the mean weight is approximately one.
The ET and FE results are obtained by averaging over 50 independent
realizations, and the error bars represent the standard deviations. For undirected ER networks, $\langle k\rangle = Np_{\rm con}$, where $p_{\rm con}$ is the connecting probability between each pair of nodes. Thus, $p_{\rm con} = \langle k\rangle / N$. For directed ER networks, $\langle k\rangle = Np_{\rm con} / 2$, yielding $p_{\rm con} = 2\langle k\rangle / N$.
}
\label{Fig.1}
\end{center}
\end{figure}
We first apply our minimum output analysis to undirected Erd\"{o}s-R\'{e}nyi (ER)
random~\cite{erd6s1960evolution} and scale-free
(SF)~\cite{barabasi1999emergence} networks and derive analytical results.
Figure~\ref{Fig.1} shows that, as the average degree
$\langle k\rangle$ ($\langle k\rangle \equiv\frac{1}{N}\sum_{i}^{N}k_{i}$,
where $k_i$ is the node degree of $i$)
is increased, $n_\text{m}$ decreases for undirected ER random networks with
identical and random link weights. For the random networks, the efficient
formula~(\ref{eq:FA}) can be further simplified.
In particular, for small values of $\langle k\rangle$, due to the isolated
nodes and the disconnected components, zero dominates the eigenvalue
spectrum of the matrix $L$~\cite{zhao2015intrinsic} where, for example,
each disconnected component generates at least one zero eigenvalue in $L$.
For large values of $\langle k\rangle$, we expect all eigenvalues
to be distinct without any dominant one. In this case, we can still choose
zero to be the eigenvalue associated with $a$ in Eq.~(\ref{eq:FA}). Taken
together, in a wide range of $\langle k\rangle$ values, the efficient
formula Eq.~(\ref{eq:FA}) holds with $a=0$.
Alternatively, the value of $n_\text{m}$ for ER networks can be theoretically
estimated using the degree distribution because of the dominance of the null
eigenvalue (see Supplemental Material S4):
\begin{equation} \label{eq:UER}
n^\text{UER}_\text{m} \approx\begin{cases}
1-\langle k\rangle/2  & \langle k\rangle\in[0, 1]\\
\frac{1}{\langle k\rangle}(f(\langle k\rangle)-f(\langle k\rangle)^2/2) & \langle k\rangle\in(1, \infty),
\end{cases}
\end{equation}
where $f(\langle k\rangle)=\sum_{k=1}^{\infty}\frac{k^{k-1}}{k!}(\langle k\rangle e^{-\langle k\rangle})^{k}$.

For undirected SF networks, $a$ in the efficient formula (\ref{eq:FA}) is the
diagonal element with the maximum number of appearances in the diagonal of
matrix $L$. In the controllability framework, the density of the driver nodes
can be calculated~\cite{zhao2015intrinsic,liu2011controllability} with the cavity
method~\cite{mezard2001bethe}. The principle can be extended to analyzing
locatability measure of SF networks in a similar manner (see Supplemental Material S5). The
analytical estimation for both ER and SF networks is in good agreement with
the results of ET and FE, as shown in Figs.~\ref{Fig.1}(a-d).
Indeed, the results indicate that, choosing $a=0$ in the efficient formula
(\ref{eq:FA}) is justified for the ER networks. For small values of
$\langle k\rangle$, zero dominates the eigenvalue spectrum, and there are
a number of messenger nodes with $n_{\rm m}>1/N$. When $\langle k\rangle$
exceeds certain value, all eigenvalues become distinct, which accounts for
the result of a single driver node with $n_{\rm m} = 1/N$. This relation
holds as $\langle k \rangle$ is increased further.

We also find that random link
weights have little effect on $n_\text{m}$ for ER networks [e.g., comparing
Fig.~\ref{Fig.1}(a) with Fig.~\ref{Fig.1}(b)], due to the fact that an ER network tends to have
many isolated components. In contrast, for SF networks, random link weights can
induce a dramatic difference from the case of identical link weights,
as shown in Fig.~\ref{Fig.1}(c) with Fig.~\ref{Fig.1}(d). Particularly, a
single messenger node is sufficient to locate sources for random link
weights with weak noise, regardless of the values of $\langle k\rangle$ and $N$. This
phenomenon can be explained based on Eq.~(\ref{eq:MMT2}), where random
link weights can be regarded as imposing perturbation to the eigenvalues
of the relevant unweighted Laplacian matrix (the locations of nonzero
elements in the two matrices are the same).
If the network has a single component, the unweighted Laplacian matrix
has only one zero eigenvalue in the spectrum. The random link weights
will shift the nonzero eigenvalues in the spectrum, making the probability
of finding two or more identical eigenvalues effectively zero. We then
expect to find one null eigenvalue and $N-1$ distinct nonzero eigenvalues
so that the entire spectrum contains eigenvalues that are all distinct. As
a result, according to Eq.~(\ref{eq:MMT2}), we have $N_\text{m}=1$ for the
undirected, single-component SF network with random link weights. A
generalization is that, for an arbitrary undirected network with random link
weights and multiple components, the value of $N_\text{m}$ is exclusively
determined by the number of components, $N_\text{c}$, i.e.,
$N_\text{m}=N_\text{c}$, due to the fact that each component contributes a
null eigenvalue. Consequently, the maximum eigenvalue degeneracy that determines
$N_\text{m}$ is equal to the number of components, $N_\text{c}$.

We now turn to directed ER and SF networks. For unidirectional
links in such a network, the average degree of the network is
$\langle k\rangle=\langle k_{\rm out}\rangle /2 =\langle k_{\rm in}\rangle /2$,
where $k_{\rm out}$ and $k_{\rm in}$ denote the out-degree and in-degree,
respectively. For directed ER networks, the FE formula is Eq.~(\ref{eq:FA})
with $a=0$. Analytical prediction of $n_\text{m}$ can be obtained based on
the FE (see Supplemental Material S4):
\begin{equation} \label{eq:DER}
n^\text{DER}_{\rm m} \approx e^{-\langle k\rangle} +
\frac{\langle k\rangle^2e^{-2\langle k\rangle}}{4}.
\end{equation}
For directed SF networks, the FE formula is still Eq.~(\ref{eq:FA}) with
$a= 0$, $-1$ or $-2$ (see Supplemental Material S4). The quantity $n_{\rm m}$ can be
theoretically predicted via (see Supplemental Material S4)
\begin{equation} \label{eq:DSF}
n^\text{DSF}_{\rm m} \approx \sum_{k=m}^{N-1}2^{-k}P(k),
\end{equation}
where $k$ is node degree and $P(k)=P(k_{\rm in}+k_{\rm out})$ is the degree
distribution. Figure~\ref{Fig.2}(e-h) show, for directed ER and SF networks, the
results of $n_{\rm m}$ from FE and analytical prediction agree well with
those from ET without any approximations.

It is noteworthy that for directed networks with random link weights,
$N_{\rm m}$ is not determined by the number of components, $N_{\rm c}$,
because there can be more than one zero in the eigenvalue spectrum of
a component, a situation that differs from that for undirected networks.
In particular, for a directed network, the matrix $L$ can have any
number of zero diagonal elements because any node without outgoing links
corresponds to such a diagonal element. According to the minimum output
analysis, there can then be any number of messenger nodes in a component.
As a result, in contrast to undirected networks with random weights, the
quantity $N_{\rm m}$ in directed networks with random link weights should be
calculated by using either Eq.~(\ref{eq:MMT}) or Eq.~(\ref{eq:FA}) for
sparse networks, not by counting the number of disconnected components.


\subsection{Source locatability of real networks}
\begin{figure}[!htb]
\begin{center}
\epsfig{figure=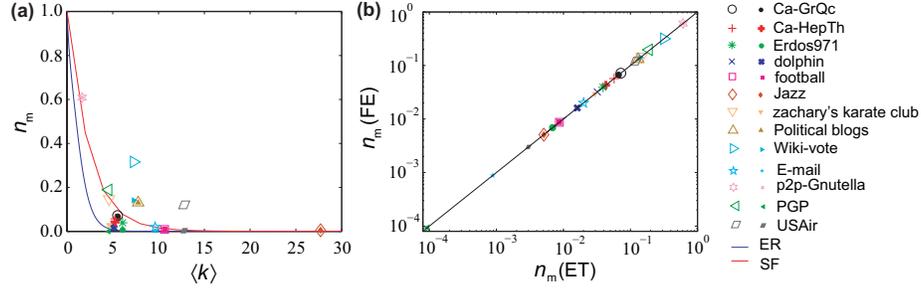,width=\linewidth}
\caption{source locatability of empirical networks.
(a) The locatability measure $n_{\rm m}$ as a function of average degree $\langle k\rangle$ for a number of real social and technological networks, on which diffusion and spreading processes may occur. (b) The locatability measure obtained by using exact theory $n_{\rm m}({\rm ET})$ [Eq.~(\ref{eq:MMT}) or Eq.~(\ref{eq:MMT2})] and obtained by using fast estimation $n_{\rm m}({\rm FE})$ [Eq.~(\ref{eq:FA})] of real networks. Here, $\langle k\rangle = \langle k_{\rm in}\rangle /2 = \langle k_{\rm out}\rangle /2$ for a directed network. Theoretical results of ER network (Eq.~(\ref{eq:UER})) and SF network with $\gamma = 3$ (Eq.~(\ref{eq:DSF})) are shown as a reference. Hollow symbols represent the results of unweighted real networks and solid symbols represent the results of real networks with random link weights selected from a uniform distribution in the range $(0,~2)$. More details of the real networks can be found in Supplemental Material S6 and Table~S1.}
\label{Fig.real}
\end{center}
\end{figure}
We also investigate the source locatability $n_{\rm m}$ for a number
empirical social and technological networks, on which diffusion or spreading
processes may occur. Because of the lack of link weights in
the real networks, we consider two typical scenarios, unweighted networks
and random weight distribution. As shown in Fig.~\ref{Fig.real}(a), $n_{\rm m}$ for an
unweighted real network is always larger than or equal to that of the
network with random weights, indicating that random link weights are
beneficial to source localization. Another feature is that sources in
the technological networks with heterogeneous degree distribution (e.g.,
Wiki-vote, p2p-Gnutella, PGP, Political blogs, USAir) are usually more
difficult to be located than the social networks with relatively
homogeneous degree distribution.

We also test the
practical feasibility of our fast estimation approaches by using the
real networks. As shown in Fig.~\ref{Fig.real}(b), we obtain a good agreement between
$n_{\rm m}({\rm ET})$ based on the exact locatability theory with
high computational complexity and $n_{\rm m}({\rm FE})$ from the fast
estimation with much higher efficiency for both unweighted and weighted
real networks with random weights. These results validate our fast
estimation approach as applied to real networks. (The characteristics of
the real networks are described in Supplemental Material S6 and Table~S1).

Combining the results of real and model networks, we discover that the
average node degree, the degree distribution and the link weight distribution
jointly determine the source locatability. In particular, sources in
networks with a homogeneous degree distribution, more connections and
random link weights are more readily to be located.



\subsection{Identification of messenger node set}
\begin{figure}[!htb]
\begin{center}
\epsfig{figure=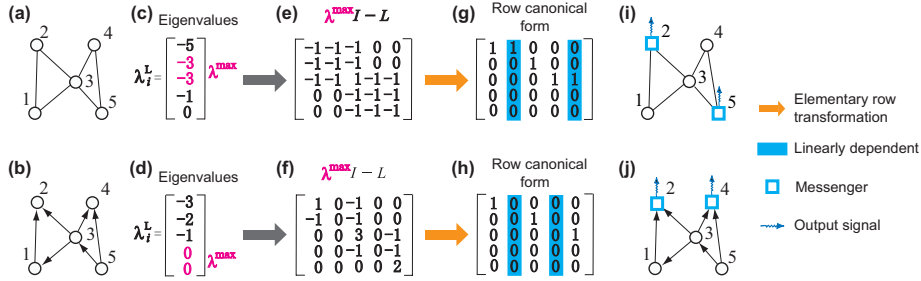,width=\linewidth}
\caption{Identification of messengers.
(a-b) Illustration of our method to identify messenger nodes for (a) a
simple undirected network and (b) a simple directed network. (c-d) Eigenvalues the undirected network in (a) and that of the directed network in (b), respectively. In (c) and (d), the eigenvalue $\lambda^{\max}$ corresponding to the maximum geometric multiplicity $\mu(\lambda^{\max})$ is highlighted in red. (e-f) Matrix $\lambda^{\max}I-L$ for the network in (a) and (b), respectively, where $\lambda^{\max}$ is highlighted. (g-h) Row canonical form of the matrix in (e) and (f) as a result of elementary row transformations, respectively. Here, linearly dependent columns in (g) and (h) are highlighted in blue. (i-j) Messenger nodes corresponding to the linearly dependent columns in the network in (a) and (b), respectively, and output signals produced by messenger nodes. For the network in (a) and (b), the
configuration of messengers is not unique as it depends on the
elementary row transformation, but the number of messengers $N_\text{m}$ is fixed and
solely determined by $\mu(\lambda^{\max})$.}
\label{Fig.2}
\end{center}
\end{figure}
We demonstrate how the $N_\text{m}$ messenger nodes can be identified using
the theory of exact observability of complex networks~\cite{yuan2013exact}.
In particular, according to the classic Popov-Belevitch-Hautus test
theory~\cite{hautus1969controllability} and our locatability theory, the
output matrix $C$ associated with the $N_\text{m}$ messenger nodes
satisfies the rank condition ${\rm rank}\begin{pmatrix}
\lambda^{\max}I-L \\
C
\end{pmatrix}=N$,
where $\lambda^{\max}$ is the eigenvalue with the maximum geometric
multiplicity $\mu(\lambda^{\max})$ of matrix $L$, i.e.,
$N-{\rm rank}(\lambda^{\max}I-L)$ reaches maximum value that is
nothing but $N_\text{m}$ (see Eq.~(\ref{eq:MMT}) and (Supplemental Material S3)).
Messenger nodes can be identified insofar as the output matrix $C$ is determined.
The computation complexity of our elementary transformation is $O(N^2(\log N)^2)$~\cite{grcar2011ordinary}.
Figure~\ref{Fig.2}(a-j) illustrate, for an undirected and a directed network, the
working of our method of identifying the messengers. For each case, we
first compute the eigenvalues $\lambda_{i}^{L}$ of the matrix $L$ and
find the eigenvalue $\lambda^{\max}$ corresponding to $\mu(\lambda^{\max})$.
We then implement elementary row transformation on $\lambda^{\max}I-L$ to obtain
its row canonical form that reveals a set of linearly-dependent columns.
The messenger nodes are nothing but the nodes corresponding to the columns
that are linearly dependent on other columns. The minimum number of messenger
nodes (linearly-dependent columns) is exactly $N_\text{m}$. Note that,
alternative configurations of the messenger nodes are possible. For example,
as shown in Fig.~\ref{Fig.2}(g), we find that columns 1 and 2, and columns 4 and 5 are
linearly correlated, requiring two messengers. As a result, there are four
equivalent combinations for the messenger nodes: (1,~4), (1,~5), (2,~4)
and (2,~5), any of which can be chosen.

\section{Source localization based on compressive sensing}
\begin{figure}[!htb]
\begin{center}
\epsfig{figure=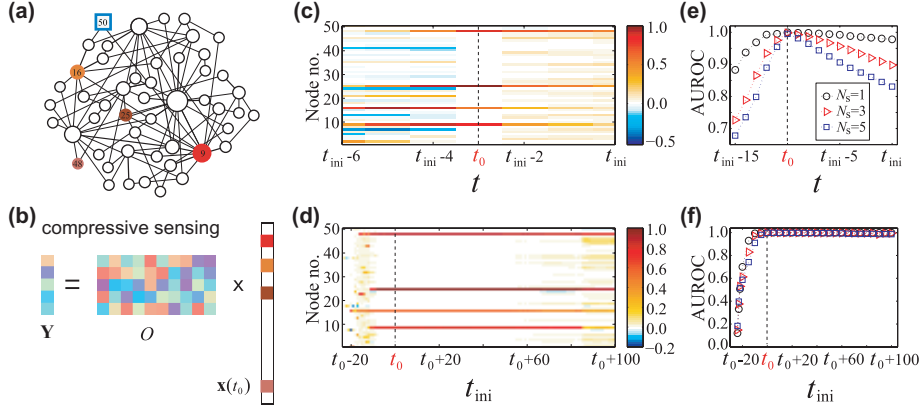,width=\linewidth}
\caption{An example of locating sources in undirected weighted
SF networks. (a) Illustration of an SF network with four sources with
colors representing the initial state values. One messenger node is
specified as a blue square. The thickness of the links represents their weight and the sizes of the nodes indicate their degrees. (b) The form of ${\bf Y}=O{\bf x}(t_0)$ and the sparse initial state vector ${\bf x}(t_0)$ to be reconstructed by using compressive sensing from relatively a small amount of data. (c) Reconstructed state $x_{i}(t)$ of each node for $t\leq t_{\rm ini}$, where the initial observation time
is $t_{\rm ini}$ $(t_{\rm ini}\geq t_{0})$. Colors represent the values of $x_{i}(t)$ with $t\leq t_{\rm ini}$. (d) Reconstructed initial state
$x_{i}(t_0)$ of each node from different initial observation time
$t_{\rm ini}$ when $t_{0}$, the true triggering time, is being
successfully inferred. Colors represent the reconstructed values of
$x_{i}(t_0)$. The colors have the same meanings as those in (a). The
four sources are randomly selected and their $x_{i}(t_0)$ values are
larger than zero. (e) AUROC as a function of $t$ ($t\leq t_{\rm ini}$)
for a fixed initial observation time $t_{\rm ini}$. (f) AUROC versus
$t$ for different initial observation time $t_{\rm ini}$ and different
number of sources ($N_{\rm s}$).
Network parameters are set as follows. Network size is $N= 50$, the average
degree is $\langle k\rangle = 4$, and the random link weights are selected
from a uniform distribution in the range $(0,~2)$. For the diffusion
dynamics, we set the diffusion parameter to be $\beta=0.05$ and the
initial state of sources in ${\bf x}(t_0)$ is randomly selected from
a uniform distribution in the range $(0.1,~1)$. To implement the source
localization process, the parameters are: noise amplitude $\sigma = 0$,
Data$= 0.5$, and the results are obtained by averaging over 300 independent
simulations.
}
\label{Fig:examp_locate}
\end{center}
\end{figure}
A result from the canonical observability theory is that, in order
to fully reconstruct ${\bf x}(t_0)$ from solutions of Eq.~(\ref{eq:main}),
at least $N$-step measurements from the messenger nodes are necessary.
However, for our localization problem, the sources are ``minority'' nodes
in the sense that the number of sources is much smaller than the network
size. In fact, the states of most nodes in the network are zero initially,
indicating that the vector ${\bf x}(t_0)$ is sparse with a large number of
zero elements. The sparsity of ${\bf x}(t_0)$ can be exploited to greatly
reduce the measurement requirement. In particular, in the CS framework
for sparse signal reconstruction~\cite{donoho2006compressed,han2015robust},
Eq.~(\ref{eq:main}) can be solved and accurate reconstruction of
${\bf x}(t_0)$ can be achieved through solutions of the following
convex-optimization problem:
\begin{equation}\label{eq:CS1}
\min\|{\bf x}(t_0)\|_{1}~~{\rm subject~~to}~~{\bf Y}=O\cdot {\bf x}(t_0),
\end{equation}
where $\|{\bf x}(t_0)\|_{1}=\sum_{i=1}^N |{\bf x}_i(t_0)|$ is the $L_1$
norm of ${\bf x}(t_0)$, ${\bf Y} \in \mathbb{R}^{qM}$,
$O \in \mathbb{R}^{qM\times N}$ and ${\bf x}(t_0) \in \mathbb{R}^{N}$.

If $O$ satisfies the restricted isometry property
(RIP)~\cite{candes2005decoding}, a full reconstruction of ${\bf x}(t_0)$
can be guaranteed theoretically through $M$-step measurements via
some standard optimization method, where $M\ll N$. For realistic
complex networks, the RIP may be violated, but because of the
linear independence of rows in matrix $O$ it is still feasible to
reconstruct ${\bf x}(t_0)$ from sparse data, where $M$ can still be
much smaller than $N$. Another advantage associated with the CS
framework lies in its robustness against noise. Especially, to obtain
the direct solution of ${\bf x}(t_0)$ is not possible when there
is measurement noise or measurements are not sufficient ($M<N$), but
the CS framework overcomes these difficulties.

A complete description of our framework to reconstruct the initial
states with unknown $t_{0}$ is described in Methods 5.2.
Here we present an example of locating diffusion sources in an SF network, as
shown in Fig.~\ref{Fig:examp_locate}. For an SF network
of a single connected component and random link weights, our minimum
output analysis gives $N_\text{m}=1$, and the single messenger node can
be selected arbitrarily. As shown in Fig.~\ref{Fig:examp_locate}(a) for an SF network with four
sources and a single messenger node.
For convenience, we define $\mbox{Data}\equiv M/N$,
i.e., the ratio of the utilized amount of measurement to the amount
required by the canonical observability theory.
Figure~\ref{Fig:examp_locate}(b) shows the form of
${\bf Y}=O{\bf x}(t_0)$, in which the initial state vector ${\bf x}(t_0)$
is to be reconstructed. Note that ${\bf x}(t_0)$ is quite sparse with four
nonzero elements corresponding to the four sources. Thus, ${\bf x}(t_0)$
can be reconstructed by using the compressive sensing from relatively a
small amount of data. Figure~\ref{Fig:examp_locate}(c) shows, for $\mbox{Data}=0.5$ and in
absence of noise, four sources and their locations as well as the
initial (triggering) time $t_{0}$ can be accurately inferred, even
though $t_{0}$ is unknown. We see that the reconstructed state
${\bf x}(t_\text{ini}-3)$ is the sparsest in the sense that it is
sparser than all the other states before and after $t_\text{ini}-3$. This
indicates that the initial time is $t_0=t_\text{ini}-3$ and
${\bf x}(t_\text{ini}-3)$ is the initial state, in which
$x_i(t_\text{ini}-3)$ with nonzero values correspond to sources.

An alternative criterion for inferring initial time $t_0$ is that ${\bf x}(t_0)$ is nonnegative but some elements in ${\bf x}(t_0-1)$ are negative. The presence of negative values in ${\bf x}(t_0-1)$ is because of the violation of physical process at time $t_0-1$. Actually, the diffusion process at $t_0-1$ does not exist, such that there is no physical solution of ${\bf x}(t_0-1)$, regardless of using any methods to solve ${\bf x}(t_0-1)$. A forced solution of ${\bf x}(t_0-1)$ will account for unreasonable values in ${\bf x}(t_0-1)$. As a result, negative values in ${\bf x}(t_0-1)$, ${\bf x}(t_0-2)$, $\cdots$ are highly possible, and offer an alternative way to the sparsity ${\bf x}(t)$ for inferring $t_0$.


In this manner, not only can we locate the sources but we can also infer
the initial states of the source nodes. As shown in Fig.~\ref{Fig:examp_locate}(c), the
reconstructed initial state values of the sources at $t=t_0$ are
in good agreement with those shown in Fig.~\ref{Fig:examp_locate}(a) (see Methods 5.2
for more details). Figure~\ref{Fig:examp_locate}(d) shows how different initial observation
time $t_{\rm ini}$ affects source localization. We find that, in the wide
range of $t_{\rm ini}$ from $t_{\rm ini}=t_0-10$ to $t_{\rm ini}=t_0+80$,
four sources can be precisely located from a small amount of data. Here,
$t_{\rm ini}<t_0$ indicates that we started to observe messenger nodes
prior to the occurrence of the diffusion event from the four sources,
which is possible because $t_0$ is unknown. If $t_{\rm ini}$ is much
earlier than $t_0$, the spreading process may not occur after $M$-step
measurements, rendering source localization impossible using any method
in principle. This accounts for the failure of our method for
$t_{\rm ini}< t_0-20$. Also, if $t_{\rm ini}$ is much
later than $t_0$, computing errors and noisy effect will be amplified
by using the CS-based optimization, leading to the inaccuracy of source
localization, e.g., $t_{\rm ini}>t_0+90$. These issues notwithstanding,
our method is quite effective for a vast range of $t_{\rm ini}$ for
multiple sources based on sparse data from a minimum number of
messenger nodes.

\begin{figure}[t]
\begin{center}
\epsfig{figure=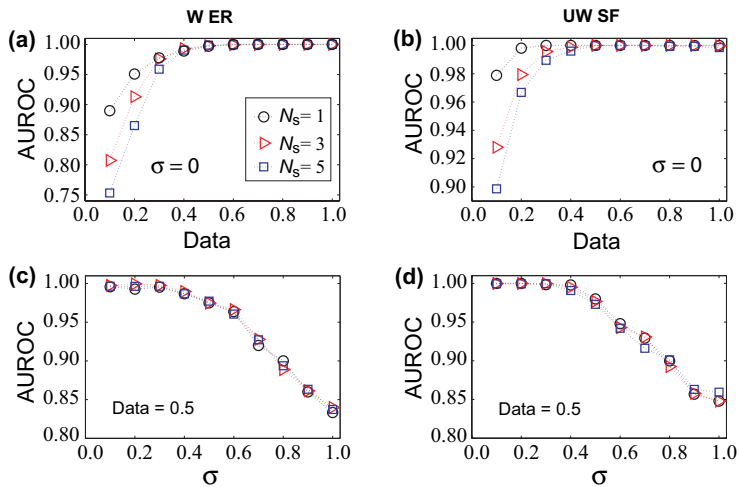,width=0.8\linewidth}
\caption{ Locatability performance in undirected ER and SF networks.
(a-d) AUROC as a function of Data for (a) weighted ER and (b) unweighted SF networks, and as a function of noise variance $\sigma$ for (c) weighted ER and (d) unweighted SF networks.
In (a) and (b), $\sigma$ is fixed at 0. In (c) and (d), Data is fixed
at 0.5. Cases with different numbers of sources, $N_\text{s}$, are
included. For a random guess, the AUROC value is 0.5. The average
degree $\langle k\rangle$ is 2 and 4 for the ER and SF networks, respectively.
We set $\beta=0.1$ for ER networks and $\beta=0.05$ for SF networks, respectively. The results are obtained by averaging over 500 independent simulations. The other parameters are the same as in Fig.~\ref{Fig:examp_locate}.}
\label{Fig.5}
\end{center}
\end{figure}
To characterize the performance of our source localization method, we use
a standard index from signal processing, the area under a receiver operating
characteristic (AUROC)~\cite{fawcett2006introduction,hanley1982meaning}.
In particular, $\mbox{AUROC}=1$ indicates the existence of a threshold
that can entirely separate the initial states ${\bf x}(t_0)$ of the sources
from other nodes in the network, giving rise to perfect localization of
sources (see Supplemental Material S7 for the detailed definition of AUROC). To give a
concrete example, we set $t_{\rm ini}=t_{0}+10$. Figure~\ref{Fig:examp_locate}(e) shows that
the value of AUROC reachs unity at $t_{\rm ini}-10$, namely $t_{0}$,
demonstrating a nearly perfect localization of sources with different
number. The highest reconstruction accuracy at $t=t_0$ corresponds to
the highest sparsity of the reconstructed state at $t_0$ in Fig.~\ref{Fig:examp_locate}(c).
For $t>t_0$, at an arbitrary time $t'$, the number of nodes with nonzero
states will be larger than the number of sources, because of the diffusion
from sources to the other nodes. Thus, one may not distinguish sources
from the other nodes based on the reconstructed ${\bf x}(t')$, accounting
for the lower values of AUROC at $t'$ compared to that at $t_0$. On the
other hand, consider an arbitrary time $t''$ with $t''<t_0$. At $t''$,
the spreading process has not occurred, and there is no causality
between the states at $t''$ and the observation. When we impose the
reconstruction on ${\bf x}(t'')$, we cannot obtain the true ${\bf x}(t'')$
with all zero elements but a virtual initial state vector with certain
errors as compared to ${\bf x}(t_0)$. The reconstruction errors will
cause more nonzero states on the basis of ${\bf x}(t_0)$, inducing a
denser state vector than ${\bf x}(t_0)$ and therefore lower values of
AUROC. The reconstruction errors also explain the fact that the value
of AUROC decreases more rapidly for $t<t_{0}$ than for $t>t_{0}$. Figure~\ref{Fig:examp_locate}(f)
shows the statistical results of Fig.~\ref{Fig:examp_locate}(d). We see that AUROC reaches unity
when the observation time $t_{\rm ini}$ is about 3 time steps ahead of
$t_{0}$, and the AUROC value is nearly unchanged as $t_{\rm ini}$ is
further increased, which is consistent with the phenomena shown in Fig.~\ref{Fig:examp_locate}(d).
(In addition, examples of locating sources in ER networks with and without
measurement noise, and in SF networks with measurement noise are presented in
Supplemental Material S8 and Fig.~S1-S3).
Here we choose the node number 50, i.e. no.50, to be the messenger. We also find the different choices of messengers don't
affect the result of the sources localization, see Supplemental Material
S8 and Fig. S4 for the details. We also investigate effects of the network size on the sources localization, and find that the Data will be smaller
for a larger network size when AUROC reaches 1, see Supplemental Material S8 and Fig. S5. This is because that the initial state ${\bf x}(t_0)$ is sparser when the network size is larger, for a certain AUROC, then the amount of data will be smaller by using CS methods.


We also systematically test the performance of our locatability framework
with respect to data requirement and robustness against noise.
We assume that measurements are contaminated by white Gaussian noise:
${\bf \hat{y}}(t)={\bf y}(t)[I+\mathcal{N}({\bf 0},\sigma^{2}I)]$, where ${\bf 0}\in\mathbb{R}^{N}$ is zero vector and
$I\in\mathbb{R}^{N\times N}$ is the identity matrix,
and $\sigma$ is the standard deviation.
The results of AUROC as a function of Data
for ER and SF networks are shown in Fig.~\ref{Fig.5}(a) and Fig.~\ref{Fig.5}(b),
respectively. In absence of noise ($\sigma=0$), even for
$\mbox{Data}=0.1$, high values of AUROC can be achieved, e.g., 0.9,
especially for SF networks. The value of AUROC exceeds 0.95 when
the amount of data is 0.3, and reaches unity for
$\mbox{Data} \geq 0.5$. The essential feature holds in presence of
noise and for arbitrary value of $N_{\rm m}$ (see Supplemental Material S9
and Fig.~S6). Another finding is that, fewer sources
(smaller values of $N_{\rm s}$) require less data, due to the fact that
a sparser ${\bf x}(t_0)$ is induced as a result of smaller $N_{\rm s}$ and
in general, the CS framework requires less amount of data to reconstruct
a sparser vector. Systematic results on noise resistance is shown in
Fig.~\ref{Fig.5}(c) and Fig.~\ref{Fig.5}(d), where we see that the AUROC
value is nearly indistinguishable across different number of sources,
$N_{\rm s}$. This is different from the results in Fig.~\ref{Fig.5}(a)
and Fig.~\ref{Fig.5}(b), and there is almost no difference between the results
from ER and SF networks. Figure~\ref{Fig.5}(c) and Fig.~\ref{Fig.5}(d) also
show that, as $\sigma$ is increased from 0 to 1, the AUROC value is
only slightly reduced (AUROC $\approx 0.85$ for $\sigma=1$), indicating
the extraordinary robustness of our locatability framework against noise.
We also study the effect of the diffusion parameter $\beta$ on source
localization with respect to different data amounts and values of the
noise variance. We find that $\beta$ has little influence on the accuracy
of source localization (see Supplemental Material S10, Fig.~S7-S9).


\section{Discussion}


We developed a framework for locating sources of diffusion or spreading dynamics
in arbitrary complex networks (directed or undirected,
weighted or unweighted) based solely on sparse measurement from a
minimum number of messenger nodes.
The key to the general framework lies in combining the controllability
theory of complex networks with the compressive sensing paradigm for
sparse signal reconstruction, both being active areas of research in
network science and engineering. Particularly, the minimum set
of messenger nodes can be identified efficiently using the minimum
output analysis based on exact controllability of complex networks and
the dual relation between controllability and observability.
The ratio of the minimum messenger nodes to the network size characterizes
the source locatability of complex networks. We find that sources in a
denser and homogeneous network are more readily to be located, which
distinguishes our work from those in the literature based on alternative
algorithms. A 
finding is that, for undirected networks
with one component, random link weights and weak noise, a single messenger node is
sufficient to locate sources at any locations in the network.
By using the data from the minimum set of messenger nodes, an approach
based on compressive sensing is offered to precisely infer the initial
time, at which the diffusion process starts, and the sources with nonzero
states initially. Because the initial state vector
to be recovered for source localization is generically sparse, compressive
sensing can be employed to locate the sources from small amounts of
measurement, making our framework robust against insufficient data and noise.
Practically, the highlights of our framework consist of the following three
features: minimum messenger nodes, sparse data requirement, and strong
noise resistance, which allow the sources of dynamical processes to
be identified accurately and efficiently.


Our approach was partially
inspired by the pioneering effort in connecting the conventional
observability theory for canonical linear dynamical systems with the compressive
sensing approach~\cite{tarfulea2011observability,dai2013observability,sanandaji2014observability}.
To our knowledge, the source locatability problem has not been tackled
in such a comprehensive way prior to our work. The minimal output analysis
based on the controllability and observability theory for complex networks
deepens our understanding of the dynamical processes on complex networks, which
finds applications, e.g., in the design and analysis of large scale sensor
networks.
Incorporating compressive sensing to uncover the sources and the original
time of diffusion represents an innovative approach to a practical problem
of significant interest but limited by finite resources for collecting
data and by measurement or background noise.
The underlying principle of the framework can potentially be applied
to solving other optimization problems in complex networks. While we
study diffusion models on time invariant complex networks, our general framework
provides significant insights into the open problem of developing source
localization methods for time variant complex networks hosting nonlinear
diffusion processes.

%
%

\section{Methods}

\subsection{The main localization formula}
The detailed form of ${\bf Y} = O \cdot {\bf x}(t_{0})$ is
\begin{eqnarray} \label{eq:11}
\begin{pmatrix}
{\bf y}(t_{0})\\
{\bf y}(t_{0}+1)\\
\vdots \\
{\bf y}(t_{0}+N-1)
\end{pmatrix}
&=&
\begin{pmatrix}
C\\
C\big[I+\beta L\big]\\
\vdots \\
C\big[I+\beta L\big]^{N-1}
\end{pmatrix}
{\bf x}(t_{0}),
\end{eqnarray}
where $N$ time steps of measurements are necessary to ensure full rank
of the observability matrix $O$. Insofar $O$ is of full rank, according
to the canonical observability theory, there exist a unique solution of
the initial states to the main localization function.

\subsection{Reconstruction of initial state ${\bf x}(t_{0})$ without
knowledge of initial time $t_{0}$}
For realistic diffusive processes on networks, the initial time $t_0$
is usually not known {\em a priori}, making inference of the initial state
${\bf x}(t_{0})$ a challenging task. Taking advantage of the sparsity
of the initial vector ${\bf x}(t_{0})$ and the underlying principle of
compressive sensing, we articulate an effective method to uncover both
${\bf x}(t_{0})$ and $t_{0}$ from limited measurements.

Say the initial observation time is $t_{\rm ini}$ $(t_{\rm ini}\geq t_{0})$.
Considering all possible $t_0$ ahead of $t_{\rm ini}$, we need to
reconstruct a series of states, i.e., ${\bf x}(t_{\rm ini})$,
${\bf x}(t_{\rm ini}-1)$, $\cdots$, ${\bf x}(t_{0}^{'})$ to ensure that
the actual $t_0$ lies in between $t_{\rm ini}$ and $t_0'$. The series of
states can be reconstructed from the uninterrupted observation
${\bf y}(t_{\rm ini}),\cdots, {\bf y}(t_{\rm ini}+N-1)$ according to the
following equations:
\begin{eqnarray} \label{eq:infer1}
\begin{pmatrix}
{\bf y}(t_{\rm ini}) \\
{\bf y}(t_{\rm ini}+1) \\
\vdots \\
{\bf y}(t_{\rm ini}+N-1)
\end{pmatrix}
&=&
\begin{pmatrix}
C\\
C\big[I+\beta L\big]\\
\vdots \\
C\big[I+\beta L\big]^{N-1}
\end{pmatrix}
{\bf x}(t_{\rm ini}), \nonumber \\
\begin{pmatrix}
{\bf y}(t_{\rm ini}) \\
{\bf y}(t_{\rm ini}+1) \\
\vdots \\
{\bf y}(t_{\rm ini}+N-1)
\end{pmatrix}
&=&
\begin{pmatrix}
C\big[I+\beta L\big] \\
C\big[I+\beta L\big]^{2} \\
\vdots \\
C\big[I+\beta L\big]^{N}
\end{pmatrix}
{\bf x}(t_{\rm ini}-1),\\
&\vdots& \nonumber \\
\begin{pmatrix}
{\bf y}(t_{\rm ini}) \\
{\bf y}(t_{\rm ini}+1) \\
\vdots \\
{\bf y}(t_{\rm ini}+N-1)
\end{pmatrix}
&=&
\begin{pmatrix}
C\big[I+\beta L\big]^{t_{\rm ini}-t_{0}^{'}} \\
C\big[I+\beta L\big]^{t_{\rm ini}-t_{0}^{'}+1} \\
\vdots \\
C\big[I+\beta L\big]^{t_{\rm ini}-t_{0}^{'}+N-1}
\end{pmatrix}
{\bf x}(t_{0}^{'}). \nonumber
\end{eqnarray}
The reconstruction process is terminated and $t_0$ can be inferred if
a sparsest state is identified, say ${\bf x}(t_{1})$, i.e., ${\bf x}(t_{1})$
is sparser than all reconstructed states at time before and after $t_1$.
Then ${\bf x}(t_{1})$ is taken as the initial state with the initial time
$t_{0}=t_{1}$.

By exploiting the natural sparsity of ${\bf x}(t)$, the CS framework for
sparse signal reconstruction allows us to reconstruct ${\bf x}(t_{\rm ini})$,
${\bf x}(t_{\rm ini}-1)$, $\cdots$, ${\bf x}(t_{0}^{'})$ iteratively
from a small amount of data, i.e., $M$-step measurements and $M < N$,
i.e., ${\bf Y} \in \mathbb{R}^{qM}$, $O \in \mathbb{R}^{qM\times N}$
and ${\bf x}(t_0') \in \mathbb{R}^{N}$.
In contrast, at least $N$-step measurements are required in the
conventional observability theory [Eq.~(\ref{eq:infer1})], where
$M$ depends on the sparsity of the state vector. In general, $M$ can be
much smaller than $N$, insofar as the number of sources $N_\text{s}$ is
much smaller than the network size $N$. According to Eqs.~(\ref{eq:CS1})
and (\ref{eq:infer1}), ${\bf x}(t_{\rm ini})$, ${\bf x}(t_{\rm ini}-1)$,
$\cdots$, ${\bf x}(t_{0}^{'})$ can be reconstructed efficiently from a
small amount of observation that is much smaller than that required in
the conventional observability theory.

\paragraph{Data accessibility.} Data can be accessed at http://sss.bnu.edu.cn/\%7Ewenxuw/data\%5fset.htm.

\paragraph{Authors¡¯ contributions.} W.X.W and Y.C.L devised the research project. Z.L.H and X.H performed numerical simulations. W.X.W, Y.C.L, Z.L.H and X.H analysed the results. W.X.W and Y.C.L wrote the paper. All authors gave final approval for publication.

\paragraph{Competing interests.} We declare we have no competing interests.

\paragraph{Funding.} W.-X.W. was supported by NSFC under Grant No.~61573064, CNNSF under
Grant No.~61074116, the Fundamental Research Funds for the
Central Universities and Beijing Nova Programme. Y.-C.L. was supported by ARO under Grant No.~W911NF-14-1-0504.

\paragraph{Acknowledgements.} We thank Mr. Zhesi Shen for valuable discussion and comments. 
We thank the reviewers (and the associate editor) for their helpful comments. In particular, the suggestion by the associate editor about the alternative method to determine $t_0$.


\end{document}